\documentclass[conference]{IEEEtran}

\ifCLASSINFOpdf
\else
\fi
\usepackage[dvips]{graphicx}
\usepackage{latexsym}
\usepackage[fleqn]{amsmath}
\usepackage{amssymb}
\usepackage{amsthm}
\usepackage{algorithm}
\usepackage{algorithmic}
\usepackage{color}

\theoremstyle{definition}

\newtheorem{definition}{Definition}

\newtheorem{lemma}{Lemma}

\newtheorem{theorem}{Theorem}

\newcommand{\X}[0]{{\mathcal X}}
\newcommand{\barX}[0]{\bar{\mathcal{X}}}
\newcommand{\F}[0]{{\mathcal F}}
\newcommand{\M}[0]{{\mathcal M}}
\newcommand{\C}[0]{{\mathcal C}}

\begin{document}

\title{Asymptotic Achievable Rate\\  of Two-Dimensional Constraint Codes \\based on Column by Column Encoding}

\author{\IEEEauthorblockN{Kazuya Hirata}
\IEEEauthorblockA{
Department of Computer Science and Engineering\\
Nagoya Institute of Technology\\
Email: 29414090@stn.nitech.ac.jp
}
\and
\IEEEauthorblockN{Tadashi Wadayama}
\IEEEauthorblockA{
Department of Computer Science and Engineering\\
Nagoya Institute of Technology\\
Email: wadayama@nitech.ac.jp
}
}

\maketitle

\begin{abstract}
In this paper, we propose a column by column encoding scheme
suitable for two-dimensional (2D) constraint codes and
derive a lower bound of its maximum achievable rate.
It is shown that the maximum achievable rate is equal to
the largest minimum degree of a subgraph of the
maximal valid pair graph. A graph theoretical analysis
to provide a lower bound of the maximum achievable rate is presented.
For several 2D-constraints such as the asymmetric and symmetric
non-isolated bit constraints, the values of the lower bound are evaluated.
\end{abstract}

\IEEEpeerreviewmaketitle

\section{Introduction}

Strong demands for non-volatile storage devices such as flash memories
have been explosively increasing due to the growing markets of mobile devices such as smart phones and SSDs.
According to the recent trend for pursuing higher areal information density, the size of a cell in a flash memory
is still shrinking. A smaller cell has less read/write reliability compared with a larger cell.
In particular, adjacent charged cells may interfere with each other
and the interference degrades quality of the read signals in some cases.
This phenomenon is called {\em inter-cell interference} \cite{Vee}\cite{J.D}.

It is known that inter-cell interference tends to occur
for a crisscross bit pattern consisting of  1-0-1 
in both vertical and horizontal directions (see Fig.\ref{fig:F1}(a)) \cite{soren}.
In \cite{Vee}, the relationship between two-dimensional (2D) written bit patterns and the quality of read signals
is investigated.
Their paper \cite{Vee} reported that, in multi-level cell flash memories (MLC) in which a cell represents four levels,
crisscross written patterns with 3-0-3, 3-1-3 and 3-2-3 in both horizontal and vertical directions
tend to cause read errors due to inter-cell interference.
Qin et al. \cite{Eitan} discusses  write-once-memory (WOM) codes
for reducing inter-cell interference.

A promising way to suppress inter-cell interference
is to use 2D constraint codes \cite{Artyom, Keren, Ido, rowbyrow} 
that forbid error-prone 2D patterns.
For example, by using an appropriate 2D constraint code,
it is possible to suppress the occurrences of crisscross pattern consisting of 1-0-1 (Fig.\ref{fig:F1}) and
the system is expected to be more interference-immune.

In general,  derivation of  the exact channel capacity
of a two-dimensional constraint channel is difficult.
The exact values of the 2D channel capacities are known only for limited 2D constraints.
In \cite{Ido}, Tal and Roth studied the 2D constraints on a crisscross area consisting of 5-cells 
and a square area consisting of 9-cells.
They showed a lower bound of the 2D channel capacity of the No Isolated Bit (NIB) constraint,  0.920862.
The NIB constraint is the constraint such that there are neither isolated single zero surrounded by ones
nor isolated single one surrounded by zeros (see Fig.\ref{fig:F1} (a) (b)).
Buzaglo et al. \cite{rowbyrow} investigated coding schemes that alleviates the inter-cell interference prone
patterns with  2D constraints in SLC and MLC flash memories,
and also evaluated their channel capacity.

In this paper, we propose  a simple column by column encoding  scheme for 2D constraints.
In the encoding process, the proposed encoder writes a column
of length $N$ according to an incoming message from a fixed size message alphabet.
The above process is repeated until an $N \times N$ 2D array is filled with the encoded columns and
the 2D array satisfies a given 2D constraint.
The column by column encoding is advantageous for practical implementation because
coding rate is constant for each column encoding process.
The paper provides a lower bound of the maximum achievable rate of  the column by column encoding 
process. 
The lower bound is based on the fact that the maximum achievable rate is equal to
the maximum value of the minimum degree of a subgraph of a certain state transition graph regarding
a given 2D constraint. By using graph theoretical techniques, we can evaluate 
the lower bound on the maximum
achievable rate. The proposed bound is applicable to wide classes of 2D constraints.

\section{Preliminaries}

In this section, we introduce definitions and notation required for the following discussion.

\subsection{Channel alphabet}
Let $\X \triangleq \{ 0, 1, \ldots, |\mathcal{X}|-1 \}$ be the channel alphabet for 2D arrays.
The $j$-th column of  $A \in \mathcal{X}^{N \times N}$ is denoted by $A(j)$.
That is, $A$ can be rewritten by
$
    A = (A(1), A(2), \ldots, A(N)).
$
The notation $[a]$ means the set of consecutive integers from $1$ to $a$, i.e., 
$[a] = \{1,2,\ldots, a\}$.

\subsection{Forbidden constraint}

In order to represent a 2D constraint, we here introduce  the {\em forbidden constraint} in this subsection.
We call  an alphabet  $\bar{\mathcal{X}} \triangleq \{ 0, 1, 2, \ldots,|\mathcal{X}|-1\} \cup \{ \ast \}$
an {\em extended channel alphabet}.
The symbol $\ast$ represent {\em don't care symbol}.

\begin{definition}{}
Assume that a matrix $A \in \X^{N \times N}$ is given.
If there exists a pair $(i,  j)  \in [N-2] \times [N-2]$ for a matrix $F \in \barX^{3 \times 3}$
satisfying
$
    A_{i + i' - 1, j + j' - 1} = F_{i', j'}
$
for any 
$
(i', j') \in \{ (i', j')  \in [3] \times [3] \mid F_{i', j'} \neq \ast \},
$
then we say that $A$ {\em contains}  $F$.
\end{definition}
Namely, if $A$ contains $F$ as a sub-matrix,
then  $A$ contains $F$.
The following definition gives a 2D constraint representing
that a given set of $3 \times 3$ matrices is excluded.

\begin{definition}{}
Assume that a set of matrices $\F \triangleq \{F_1, F_2, \ldots, F_U \}$ is given where
$F_i \in \barX^{3 \times 3}$.
If a matrix $A \in \X^{N \times N}$ does not contain any matrix in $\F$, then
$A$ is said to be a $\F$-constrained matrix.
We also call the set $\F$  a {\it forbidden constraint}.
\end{definition}

\begin{figure}[tb]
  \begin{center}
    \includegraphics[width = 50mm]{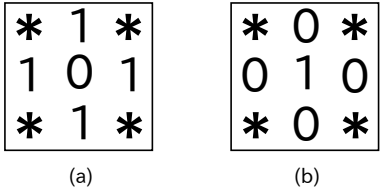}
    \caption{Forbidden matrices (a), (b): if $\F$ contains only the matrix (a), then the constraint is said to be the asymmetric NIB
    constraint. If $\F$ contains both (a) and (b) as forbidden matrices, then the constraint is said to be the symmetric NIB constraint in this paper.}
    \label{fig:F1}
  \end{center}
  \end{figure}

For a given forbidden constraint $\F$,
a set of $\F$-constrained matrices are called a $\mathcal{F}$-forbidden 2D constraint code.
In the following sections, we will discuss an encoding scheme for
$\mathcal{F}$-forbidden 2D constraint codes.
Figure \ref{fig:F1} presents examples of forbidden constraints; this constraints define
the asymmetric and the symmetric NIB constraints.

 \subsection{Column by column encoding scheme}

Let $\M \triangleq [M]$ be a message alphabet.
The column by column encoding function is defined as follows.
\begin{definition}{}
Let $\F$ be a forbidden constraint.
A function $E: \mathcal{M} \times \mathcal{X}^N \times \mathcal{X}^N \to  \mathcal{X}^N$
is called a {\em column by column encoding function} if
there exists a pair of the initial columns
\begin{eqnarray}\label{initial}
 X(-1) \in \mathcal{X}^N, X(0) \in \mathcal{X}^N,
\end{eqnarray}
 satisfying that $X = (X(1), X(2), \ldots, X(N))$ is a $\F$-constrained matrix
 for any messages $m_i \in \mathcal{M} (i \in [N])$ where
$X(i)$ is given by the following recursion:
\begin{eqnarray}\label{enc_seq}
  X(i) = E(m_i, X(i-2), X(i-1)),\quad i \in [N].
\end{eqnarray}
\end{definition}
The encoding function generates a column of length $N$ according to the message $m_i$ for each time index $i \in [N]$.
The final output $(X(1), X(2), \ldots X(N))$ can be seen as a codeword of
a $\mathcal{F}$-forbidden 2D constraint code.
It should be remarked that the message alphabet is the same for any index $i \in [N]$.
This means that the encoding rate is time invariant in an encoding process.

The following definition provides the definition of the decoding function.
\begin{definition}{}
Let $\F$ be a forbidden constraint.
Assume that a column by column encoding function $E$ is given.
For any 2D codeword
 $
    X = (X(1), X(2), \ldots, X(N))
 $
corresponding to the message symbols $m_i \in \M (i \in [N])$
generated by the encoding function $E$,
if a function $D: \X^N \times \X^N \times \X^N \to \M$ satisfies
  \begin{eqnarray}
    m_i = D(X(i-2), X(i-1), X(i)), \quad i \in [N]
  \end{eqnarray}
for any $i \in [N]$, then
$D$ is called a {\em decoding function} corresponding to the encoding function $E$.
It assumed that $D$ knows  the initial values in $(\ref{initial})$.
\end{definition}
The definition of the decoding function ensures that a set of encoded message symbols
can be perfectly recovered from a  2D array encoded by the encoding function $E$.

\subsection{Maximum achievable rate}

In this subsection, the achievable rate of 2D $\mathcal{F}$-forbidden constraint codes
is discussed.
A $\mathcal{F}$-forbidden 2D constraint code 
is defined by  a column by column encoding function $E$ as follows.
\begin{definition}{}
Let $E$ be a column by column encoding function.
The set of all the 2D coded arrays obtained by (\ref{enc_seq})
is denoted by $\C(\F, E)$.
The coding rate of $\C(\F, E)$ is defined by
$
    R(\F, E) \triangleq {\log_2|C(\F, E)|}/{(N^2\log_2|\X|)}.	
$
\end{definition}

An achievable rate can be regarded as coding rate achievable by using
a pair of a column by column encoding function $E$ and the corresponding decoding function $D$.
\begin{definition}{}
For a given positive real number $R$, a forbidden constraint $\F$, and a positive integer  $N$,
if there exists a column by column encoding function $E$
which satisfies
$
    R \leq R(\F, E)
$
exists, $R$ is said to be {\em achievable}.
\end{definition}

The supremum  of the achievable rate in the asymptotic situation
$
R^*(\mathcal{F}) \triangleq \limsup_{N \to \infty} \{R | R \ {\rm is \ achievable}\}
$
is referred to as  the {\em maximum achievable rate} of the column by column encoding.

\section{Valid pair graph and its minimum degree}


\subsection{Valid pair graph}

\begin{figure}[tb]
\centering
\includegraphics[width = 50mm]{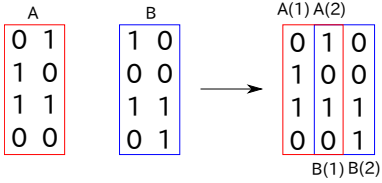}
\caption{An example of a valid pair.  The channel alphabet is $\X=\{0,1\}$.
The symmetric NIB constraint is assumed.  }\label{fig:const_pair}
\end{figure}

In the following analysis,
a class of state transition graphs corresponding to a forbidden constraint $\F$ plays an important role.
Before introducing the state transition graph,
we first define the {\em valid pair} as follows.
\begin{definition}{} \label{validpair}
Let  $\mathcal{F}$ be a forbidden constraint.
Two matrices $A \triangleq (A(1), A(2)) \in \X^{N \times 2}$
and $B \triangleq (B(1), B(2)) \in \X^{N \times 2}$ are also given.
The pair $(A, B)$ is said to be a {\em valid pair} if
$A(2)=B(1)$  and
$
  (A(1), A(2), B(2)) \in \mathcal{X}^{N \times 3}
$
does not contain any matrix in $\mathcal{F}$.
\end{definition}
Figure \ref{fig:const_pair} shows a valid pair of the symmetric NIB constraint.

\begin{figure}[tb]
\centering
\includegraphics[width = 50mm]{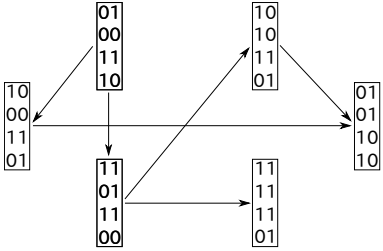}
\caption{An example of the valid pair graph with $N = 4$ and $\X = \{0,1\}$.
The symmetric NIB constraint is assumed.
A direct edge connects two nodes that are a valid pair}
\label{fig:const_pair_graph}
\end{figure}

The {\em valid pair graph} defined below is a state transition graph that indicates allowable state transitions
in a column by column encoding function for a forbidden constraint $\F$.
\begin{definition}{}
Assume that a directed graph $G = (V, E)$ is given where
any node in  $V$  belongs to $\X^{N \times 2}$.
If the edge set is given by
\[
    E = \{ (A, B) \in V \times V | (A, B) \ {\rm is \ a \ valid \ pair}\},
\]
we call $G$ a valid pair graph.
\end{definition}
Figure \ref{fig:const_pair_graph} presents an example of a valid pair graph with $N= 4$ for
the symmetric NIB constraint.

It should be noted that a valid pair is not symmetric in general, namely,
$(A, B)$  does not necessarily imply that $(B,A)$ is a valid pair.
If a forbidden constraint $\F$ makes any valid pair symmetric, i.e., a valid pair $(A, B)$ implies that $(B, A)$ is valid,
the forbidden constraint is said to be {\em symmetric}.
In the following discussion, we focus only on symmetric forbidden constraints.
In such a case,  a valid pair graph $G = (V, E)$ satisfies $(u, v) \in E \Leftrightarrow (v, u) \in E$ for any $(u, v) \in V \times V$.
Namely, the valid pair graph  $G$ becomes an undirected graph.

\subsection{Encoding function derived from valid pair graph}

Given a valid pair graph, a column by column encoding function and its decoding function
are naturally defined.
In this subsection, we will present such column by column encoding function.

Before going into the detail of the encoding function, we here prepare notation
regarding valid pair graphs. For a valid pair graph $G = (V, E)$,  the degree of a node $v \in V$
is denoted by $d(v)$ and the minimum degree of $G$ by $\delta(G)$.
The set of adjacent nodes of $v$ is denoted by
$
  \partial(v) \triangleq \{ u \in V | (u, v) \in E\}.
$
The matrix corresponding to the node $v \in V$ is expressed as $(l(v), r(v))$ where
$l(v), r(v) \in \mathcal{X}^N$.

Let $G = (V, E)$ be a valid pair graph.
For any node $v$, we prepare a bijective map $\phi_v(v \in V)$:
$
  \phi_v : [d(v)] \to \partial(v).
$
This bijective map is called an {\em encoding label map},
which connects a state transition and an incoming message in encoding processes.
A set of encoding label maps is summarized as $\phi \triangleq \{ \phi_v\}_{v \in V}$.
We call a pair of a valid pair graph $G = (V, E)$ and its encoding label map $\phi$
a {\em labeled valid pair graph}.

The following definition provides the definition  of
a column by column encoding function induced from a labeled
valid pair graph.
\begin{definition}
	Let $\F$ be a forbidden constraint.
  Assume that a labeled valid pair graph $H = ((V, E), \phi)$ is given for $\F$.
  Let $|\M| = \delta(H)$ where $\delta(H) = \delta(V, E)$.
  For $m \in \M, a, b \in \X^N$,
  we define a column by column encoding function by
  $
    E_H(m, a, b) \triangleq r(\phi_v(m)),
  $
   if there exists $v \in V$ satisfying $l(v) = a, r(v) = b$.
   Otherwise, the value of $E_H(m, a, b)$ is undefined.
  The initial value of encoding process $(\ref{enc_seq})$
  of the encoding function $E_H$
  is arbitrarily selected as $v_{init} \in V$.
  We set $X(-1) = l(v_{init}), X(0) = r(v_{init})$.
\end{definition}
From the definition of the encoding function,
It is clear that an encoded sequence
$X = (X(1), X(2), \ldots, X(N))$ generated by the encoding function $E_H$
does not contain any forbidden matrix in $\F$
for any $m_1, m_2, \ldots, m_N (m_i \in \M)$.

The next definition gives decoding function corresponding to
the encoding function $E_H$.
\begin{definition}
	Let $\F$ be a forbidden constraint.
  Assume that a labeled valid pair graph $H = ((V, E), \phi)$ for $\F$ is given.
  If there exists $v \in V$ and $m \in \M$ that satisfies
  $
    l(v) = a, \quad r(v) = b,\quad  r(\phi_v(m)) = c,
  $
then the decoding function $D_H$ is defined by
$
  D_H(a, b, c) \triangleq m.
$
The initial values are given by $X(-1) = l(v_{init}), X(0) = r(v_{init})$.
\end{definition}
It is easy to check the function $D_H$ is the decoding function corresponding to $E_H$.
We saw that constructing a labeled valid pair graph for $\F$ is equivalent to designing a pair of
encoding and decoding functions.

The coding rate of the 2D constraint code $\C(\F, E_H)$ defined by the column by column encoding function $E_H$ is given by
\begin{equation}
  R(\mathcal{F}, E_H) =\frac{\log_2 |\C(\F, E_H)|}{N^2 \log_2|\X|}
  =\frac{\log_2(\delta(H))}{N\log_2|\X|}
\end{equation}
because $|\M|$ is set to $\delta(H)$.
Namely, the rate of 2D constraint code $\C(\F, E_H)$ {\em is determined by the minimum degree of} $H$.

\section{Valid pair graph with large minimum degree}

For a given forbidden constraint $\F$, it is desirable to find a valid pair graph with large minimum degree
because it leads to higher coding rate.  In this section, we discuss how to find a valid pair graph with a large
minimum degree.

\subsection{Maximization of minimum degree}

\begin{definition}
  Let  $\F$ be a forbidden constraint.
  Let $V$ be the set of all the matrices in $\mathcal{X}^{N \times 2}$.
  The valid pair graph $G^*_N(\mathcal{F})=(V,E)$ is called the {\em maximal valid pair graph} for $\F$.
\end{definition}

It is evident that any induced subgraph $G$ of the maximal valid pair graph $G^*_N(\mathcal{F})$
is a valid pair graph.
From this fact,  we have the following theorem.
\begin{theorem}
  For a given forbidden constraint $\mathcal{F}$ and a positive integer $N$, the equality
\[
    \max\{ R \in \mathbb{R} | R \ {\rm is \ achievable}\} = R_N(\mathcal{F})
\]
    holds where the rate $R_N(\mathcal{F})$ is defined by
  \begin{eqnarray}
    R_N(\mathcal{F}) \triangleq \frac{\log_2 {\rm max}_{G \subseteq {G}^*_N(\mathcal{F})}\delta(G)}{{N\log_2|\mathcal{X}|}}.
  \end{eqnarray}
\end{theorem}
\noindent
The Proof is omitted due to the limitation of space.

This theorem claim that we need to find a subgraph of ${G}^*_N(\mathcal{F})$ with the largest minimum degree to
construct the optimal 2D constraint code with respect to coding rate.
Based on the above theorem, the maximum achievable rate for the column by column encoding is characterized as
\begin{eqnarray}\label{capacity}
  R^*(\mathcal{F}) =	\limsup_{N \to \infty}\frac{1}{N\log_2|\mathcal{X}|}\log_2 {\rm max}_{G \subseteq {G}^*_N(\mathcal{F})}\delta(G).
\end{eqnarray}

\subsection{Graph Pruning}

In the previous subsection, we saw  that it is necessary to select an induced subgraph from ${G}^*_N(\mathcal{F})$
to maximize the minimum degree.
The problem to find a subgraph with the largest minimum degree is a computationally intractable problem.
Instead of finding the optimal subgraph,
we will discuss the graph pruning method,
which is a method for finding a subgraph with the minimum degree larger than or equal to the {\em density} of the graph.

For an undirected  simple graph, we define the density of the graph $G$ by
$
  \epsilon(G) \triangleq {|E|}/{|V|}.
$
The following lemma states that
 the minimum degree of an induced subgraph $G'$ obtained from pruning process
is guaranteed to be larger than or equal to the density of the original graph.
\begin{lemma}  \cite{R}\label{pruning}
  An undirected simple graph $G = (V, E)$ with one or more edges has an induced subgraph $G' \subseteq G$ that satisfies
$
    \delta(G') \ge \epsilon(G).
$
\end{lemma}
\noindent

It should be remarked that $G'$ in this lemma is not necessarily equal to the optimal subgraph in terms of the minimum degree.
However, we can use the density of the original graph as a lower bound of the optimal minimum degree.

\subsection{Asymptotic density}

\begin{definition}{}
  Let $\tilde{G}^{*}_N(\mathcal{F})$ be the undirected simple graph obtained from the maximal valid pair graph $G^*_N(\mathcal{F})$
  by removing the self loops.
  We define the {\em asymptotic growth rate of the density} for $\tilde{G}^*_N(\mathcal{F})$ by
 \begin{eqnarray}
   \alpha(\mathcal{F}) \triangleq \lim_{N \to \infty} \frac{1}{N} \log_2 \epsilon(\tilde{G}^*_N(\mathcal{F})).
 \end{eqnarray}
\end{definition}

From the definition of the asymptotic growth rate of the density, we immediately have the following lamma.
\begin{lemma} \label{lowerbound}
  Assume that a forbidden constraint $\mathcal{F}$ is given. The inequality
  \begin{eqnarray}
    \lim_{N \to \infty} {\rm max}_{G \subseteq \tilde{G}^*_N(\mathcal{F})}
    \frac 1 N \log_2\delta(G) \ge \alpha(\mathcal{F})
  \end{eqnarray}
  holds.
\end{lemma}
\noindent
(Proof) The inequality is a direct consequence of Lemma \ref{pruning}.

From this lemma,  it is straightforward to derive a lower bound on the maximum achievable rate.
\begin{theorem}\label{alpha}
  Let  $\F$ be a forbidden constraint.
  The following inequality
$
  R^*(\mathcal{F}) \ge {\alpha(\mathcal{F})}/{\log_2 |\mathcal{X}|}\label{C}
$
holds.
\end{theorem}
\noindent
(Proof)
It is obvious from (\ref{capacity}) and lemma \ref{lowerbound}.

\subsection{Evaluation of asymptotic growth rate}

According to the discussion in the previous subsection,
it can be seen that evaluation of the asymptotic growth rate $\alpha(\mathcal{F})$ is needed to have
a lower bound on the maximum achievable rate.
In this subsection, for given forbidden constraint, we will give a method for evaluating
the asymptotic growth rate of the graph density.

The density of the simple graph $\tilde G^*_N(\mathcal{F})$ is given by
\begin{eqnarray}
  \epsilon(\tilde{G}^*_N(\mathcal{F})) = \frac{|E(G^*_N(\mathcal{F}))| - K }
  {|\mathcal{X}|^{2N}},
\end{eqnarray}
where $K$ represents the number of self loops in $G^*_N(\mathcal{F})$.
If $K = o(|E(G^*_N(\mathcal{F}))|)$ holds, then we have
\begin{eqnarray} \nonumber
  \alpha(\mathcal{F}) &=&
  \lim_{N \to \infty} \frac{1}{N}\log_2\epsilon(\tilde{G}^*_N(\mathcal{F})) \\ \nonumber
  &=& \lim_{N \to \infty} \frac{1}{N}
  \log_2 \left(|E(G^*_N(\mathcal{F}))| - K \right) - 2\log_2|\mathcal{X}| \\ \label{count}
  &=& \lim_{N \to \infty} \frac{1}{N}
  \left(\log_2 |E(G^*_N(\mathcal{F}))| \right) - 2\log_2|\mathcal{X}|.
\end{eqnarray}
The problem to evaluate $\alpha(\mathcal{F})$ is now reduced to an evaluation problem for the quantity
$\lim_{N \to \infty} (1/N) \left(\log_2 |E(G^*_N(\mathcal{F}))| \right)$, i.e, asymptotic
edge counting problem for the maximal valid pair graph $G^*_N(\mathcal{F})$.

\subsection{Counting of edges in maximum valid pair graph}

In order to evaluate the asymptotic growth rate $\alpha(\mathcal{F})$, we need to count
the number of edges in the maximum valid pair graph and to know the asymptotic behavior of it.
In this section, we will describe a method to evaluate the number of edges of the maximum valid pair graph.

Let the total number of the valid pair in the maximum valid pair graph $G^*_N(\mathcal{F})$ be $L_N(\mathcal{F})$.
The total number of edges $|E(G^*_N(\mathcal{F})|$ is then given by
$
  |E(G^*_N(\mathcal{F})| = {L_N(\mathcal{F})}/{2}
$
where the factor 2 compensates the double count of edges because $G^*_N(\mathcal{F})$ is undirected graph.
In the following discussion, we will count the number of the valid pairs $L_N(\mathcal{F})$
by using a {\em counting graph}.

Assume that a forbidden constraint $\mathcal{F}$ is given.
Let $A$ and $B$ be matrices in $\X^{2 \times 3}$, respectively.
The $i$-th row of $A$ (resp. $B$) is denoted by $A[i]$ (rep. $B[i]$).
This means  that $A$ and $B$ can be represented by
\begin{equation}
A = \left(
\begin{array}{c}
  A[1] \\
  A[2] \\
\end{array}
\right), \quad
B = \left(
\begin{array}{c}
  B[1] \\
  B[2] \\
\end{array}
\right),
\end{equation}
respectively.
Although it is an abuse of notation, we call $A$ and $B$ are valid pair if $A[2] = B[1]$  and
\begin{equation}
\left(
  \begin{array}{c}
    A[1] \\
    A[2] \\
    B[2]
  \end{array}
  \right) \in \X^{3 \times 3}
  \end{equation}
does not contain any matrix in $\F$.
The counting graph $G(\F) = (\X^{2 \times 3}, E)$ is an undirected
graph such that an edge connects two nodes in a valid pair.
From this definition of the counting graph,
it is clear that a path of length $N$ on the counting graph
corresponds to a valid pair defined in (\ref{validpair}).
Furthermore, any valid pair of length $N$ can be found in the counting graph as a path of length $N$ .
This means that counting the number of paths of length $N$ in the counting graph is
equivalent to counting the number of valid pairs.
Thus, in order to evaluate the asymptotic growth rate of the total number of paths in counting graph,
we can use a standard technique, i.e,  the growth rate can be evaluated
from the maximum positive eigenvalue of the adjacent matrix of the counting graph.

\section{Evaluation of lower bounds}

In this section, we will evaluate lower bounds of the maximum achievable rate
for several forbidden constraints based on the method presented
in the previous section.

\subsection{Asymmetric NIB constraint}
We construct the counting graph in order to count the number of edges in the valid pair graph.
Since the counting graph includes $2^6$ nodes, the adjacent matrix of the counting graph becomes 
a $2^6 \times 2^6$ integer matrix.
We can easily evaluate the maximum positive eigenvalue of the adjacent matrix, which is
$
  \lambda_{max} = 7.750.
$
The maximum eigenvalue can be used to evaluate $\alpha(\mathcal{F})$ in the following way:
\begin{eqnarray}
  \alpha(\F ) &=& \lim_{N \to \infty} \frac{1}{N} \log_2 \epsilon(G_N(F)) \nonumber \\
  &=& \lim_{N \to \infty} \frac{1}{N} (\log_2 L_N(\mathcal{F}) - 1 - 2N)  \nonumber \\
  &=& \log_2 \lambda_{max} - 2 \cong 0.954.
\end{eqnarray}
From the bound $R^*(\mathcal{F}) \ge {\alpha(\mathcal{F})}/{(\log_2 |\X|)}$ in (\ref{C}),
we obtain
\begin{eqnarray}
  R^*(\mathcal{F}) \ge \frac{\alpha(\mathcal{F})}{\log_2 |\mathcal{X}|} = \frac{\alpha(\mathcal{F})}{\log_2 2} = 0.954
\end{eqnarray}
as a lower bound of the maximum achievable rate for the asymmetric NIB constraint.

\subsection{Symmetric NIB constraint}
In the case of the symmetric NIB constraint (see Fig.\ref{fig:F1}),
we have
$
  \alpha(\mathcal{F}) \cong 0.861,
$
and thus obtain
\[
  R^*(\mathcal{F}) \ge {\alpha(\mathcal{F})}/{\log_2 2} \cong 0.861
\]
as a lower bound of the maximum achievable rate.
The value of the 2D channel capacity of this constraint is known to be 0.923 \cite{Ido}.
The lower bound 0.861 is fairly smaller than the value of the capacity.
This is because we impose additional constraints, i.e., the column by column encoding,
this may cause rate loss.

\subsection{Quaternary ICI-constraint}
Here, we derive the lower bound for the {\it quaternary ICI-constraint}.
The quaternary ICI-constraint is a constraint forbidding 3-0-3, 3-1-3 and 3-2-3 in both horizontal and vertical directions \cite{Vee}.
We can construct the counting graph and have
$
\alpha(\mathcal{F}) \cong 1.996.
$
We thus obtain
\[
R^*(\mathcal{F}) \ge \frac{\alpha(\mathcal{F})}{\log_2 4} \cong 0.998
\]
as a lower bound of the maximum achievable rate.

\section{Conclusive summary}

In this paper, we proposed a column by column encoding scheme for the 2D-constraint
imposed by a forbidden constraint $\F$ and then showed a graph theoretical analysis
for asymptotic achievable rate.
It is shown that a lower bound for the maximum achievable rate is obtained
from the maximum positive eigenvalue of the adjacency matrix of the counting graph.
For several forbidden constraints, we calculated the values of the lower bound.
In the case of binary channel alphabet, the lower bounds for the asymmetric and symmetric NIB constraints
are 0.954 and 0.861, respectively.
In the case of the quaternary channel alphabet, the lower bounds for the quaternary ICI-constraint is 0.998.

\section*{Acknowledgment}
This work was supported by JSPS Grant-in-Aid for Scientific Research  Grant Number 16K14267.


\end{document}